\documentclass[runningheads]{llncs}
\usepackage{graphicx}
\usepackage{graphicx}
\usepackage{comment}
\usepackage{siunitx}
\usepackage{subfig}
\usepackage{makecell}
\usepackage{amsmath,amssymb} % define this before the line numbering.
\usepackage{color}
\usepackage{amsmath}

\usepackage{paralist, tabularx}
\newcommand{\numc}[1]{\num[group-separator={,}]{#1}}

\begin{document}
\title{Active MR $k$-space Sampling with Reinforcement Learning}
\author{Luis Pineda\inst{1} \and
Sumana Basu\inst{2} \and
Adriana Romero \inst{1} \and
Roberto Calandra \inst{1} \and
Michal Drozdzal \inst{1}
}
%index{Pineda, Luis}
%index{Basu, Sumana}
%index{Romero, Adriana}
%index{Calandra, Roberto}
%index{Drozdzal, Michal}
% %
\authorrunning{L. Pineda et al.}
\institute{Facebook AI Research \and
 McGill University}
% \email{lep@fb.com}
% % \url{http://www.springer.com/gp/computer-science/lncs} \and
% % ABC Institute, Rupert-Karls-University Heidelberg, Heidelberg, Germany\\
% % \email{\{abc,lncs\}@uni-heidelberg.de}}
%
\maketitle              % typeset the header of the contribution

%%%%%%%%% ABSTRACT
\begin{abstract}
Deep learning approaches have recently shown great promise in accelerating magnetic resonance image (MRI) acquisition. The majority of existing work have focused on designing better reconstruction models given a pre-determined acquisition trajectory, ignoring the question of trajectory optimization. In this paper, we focus on learning acquisition trajectories given a fixed image reconstruction model. We formulate the problem as a sequential decision process and propose the use of reinforcement learning to solve it. Experiments on a large scale public MRI dataset of knees show that our proposed models significantly outperform the state-of-the-art in active MRI acquisition, over a large range of acceleration factors.


\end{abstract}

%%%%%%%%% BODY TEXT
\section{Introduction}
Magnetic resonance imaging (MRI) is a powerful imaging technique used for medical diagnosis. Its advantages over other imaging modalities, such as computational tomography, are its superior image quality and zero radiation exposure. 
Unfortunately, MRI acquisition is slow (taking up to an hour) resulting in patient discomfort and in artifacts due to patient motion.

MRI scanners sequentially acquire $k$-space measurements, collecting raw data from which an image is reconstructed (e.g., through an inverse Fourier transform). A common way to accelerate MRI acquisition is to collect less measurements and reconstruct the image using a partially observed $k$-space. 
Since this results in image blur or aliasing artifacts, traditional techniques enhance the image reconstruction using regularized iterative optimization techniques such as compressed sensing \cite{Lustig2007}. 
More recently, the inception of large scale MRI reconstruction datasets, such as \cite{zbontar2018fastMRI}, have enabled the successful use of deep learning approaches to MRI reconstruction \cite{Wang2016Accelerating,hammernik17,Schlemper2017DeepCascade,Chen2018Variable,Zhu2018Image,Zhang2018,Lonning2019Recurrent,WANG2020}. 
However, these methods focus on designing models that improve image reconstruction quality for a \emph{fixed} acceleration factor and set of measurements.

In this paper, we consider the problem of optimizing the sequence of \textit{k}-space measurements (i.e., trajectories) to reduce the number of measurements taken.
Previous research on optimizing \textit{k}-space measurement trajectories is extensive and include Compressed Sensing-based techniques \cite{seeger2010optimization,ravishankar2011adaptive,zhang2014energy,gozcu2018learning}, SVD basis techniques \cite{zientara1994dynamically,panych1996implementation,zientara1998applicability}, and region-of-interest techniques \cite{yoo1999real}.  
However, all these solutions work with \emph{fixed trajectories} at inference time. 
Only recently, on-the-fly acquisition trajectory optimization methods for deep-learning-based MRI reconstruction have emerged in the literature~\cite{jin2019self,Zhang_2019_CVPR}. 
On the one hand, \cite{jin2019self} showed that jointly optimizing acquisition trajectory and reconstruction model can lead to slight increase in image quality for a \emph{fixed} acceleration factor with subject-specific acquisition trajectories. On the other hand,~\cite{Zhang_2019_CVPR} introduced \emph{active} MRI acquisition, where both acceleration factor and acquisition trajectory are subject-specific. 
Their proposed approach performs trajectory optimization over a \emph{full range} of possible accelerations but makes a myopic approximation during training that ignores the sequentiality of $k$-space acquisition.

In contrast, we focus on Cartesian sampling trajectories and expand the formulation of \cite{Zhang_2019_CVPR}. 
More precisely, we specify the active MRI acquisition problem as a Partially Observable Markov Decision Process (POMDP)~\cite{sondik1971optimal,kaelbling1998planning}, and propose the use of deep reinforcement learning~\cite{mnih2013playing} to solve it.\footnote{For a comprehensive overview of reinforcement learning applied to healthcare data and medical imaging please refer to \cite{yu2019reinforcement}.}
Our approach, by formulation, optimizes the reconstruction over the whole range of acceleration factors while considering the sequential nature of the acquisition process -- future scans and reconstructions are used to determine the next measurement to take. 
We evaluate our approach on a large scale single-coil knee dataset~\cite{zbontar2018fastMRI}\footnote{Data used in the preparation of this article were obtained from the NYU fastMRI Initiative database (fastmri.med.nyu.edu)~\cite{zbontar2018fastMRI}. As such, NYU fastMRI investigators provided data but did not participate in analysis or writing of this report.  A listing of NYU fastMRI investigators, subject to updates, can be found at: fastmri.med.nyu.edu.  The primary goal of fastMRI is to test whether machine learning can aid in the reconstruction of medical images.} and show that it outperforms common acquisition heuristics as well as the myopic approach of \cite{Zhang_2019_CVPR}. 
Our contributions are: 
1) formulating active MRI acquisition as a POMDP; 
2) showing state-of-the-art results in active MRI acquisition on a large scale single-coil knee dataset; 
3) performing an in-depth analysis of the learned trajectories. 
The code to reproduce our experiments is available at: \url{https://github.com/facebookresearch/active-mri-acquisition}.

\section{Background}
\label{sec:background}
\textbf{Partially Observable Markov Decision Processes (POMDP)} A POMDP is a highly expressive model for formulating problems that involve sequential decisions~\cite{kaelbling1998planning,puterman1990markov,sondik1971optimal}. A solution to a POMDP is a \emph{policy} that maps history of observations (in our case, the $k$-space measurements) to actions (in our case, an index of $k$-space measurement to acquire). A POMDP is formally defined by: (1) a set of {\it states,} summarizing all the relevant information to determine what happens next;
% the set of possible conditions for the process; 
(2) a set of {\it observations,} containing indirect information about the underlying states, which are themselves not directly observable; (3) a set of {\it actions,} that cause the transition between states;
%the set of possible controls that can be passed to the system to modify its state; 
(4) a {\it transition function,} specifying a probability distribution for moving from one state to another, after taking an action; (5) an {\it emission function,} defining the probability of obtaining each possible observation after taking an action in some state; (6) a {\it reward function,} specifying the numeric feedback obtained when moving from one state to another; and (7) and a \emph{discount factor,} defining the impact of immediate versus more distant rewards. %Note that when states are fully observable a POMDP is reduced to a Markov Decision Process (MDP)~\cite{puterman1990markov}. 

\textbf{Double Deep Q-Networks (DDQN)} (DDQN)~\cite{van2016deep} is a state-of-the-art deep reinforcement learning method for solving high-dimensional POMDPs.
Since active MRI acquisition involves discrete actions (e.g., indexes of $k$-space measurements to acquire), we focus on the Double DQN (\textsc{ddqn})~\cite{van2016deep} algorithm, given its training stability and recent success. %a relatively simple algorithm with good training stability. 
In \textsc{ddqn}, policies are not explicitly modelled. Instead, a \emph{value network} is used to predict the \emph{value} of each possible action---defined as the expectation of the future cumulative reward. A policy is then recovered by greedily choosing actions with maximal estimated value.
%\footnote{In the literature, the utility is referred to as a \emph{$Q$-value} and the value network is referred to as a \emph{$Q$-network}.} 
%  where target values are computed as bootstrapped estimates of the estimated value.'
The value prediction problem is posed as a supervised regression problem, trained to minimize the temporal difference error~\cite{sutton1988learning} over data sampled from a \emph{replay memory buffer}, which contains tuples of states, actions and rewards obtained by running the learned policy in an exploratory way. The target values used to update the value network are computed as bootstrapped estimates of the estimated value. Further details can be found in~\cite{van2016deep}. 

\section{Learning Active MR $k$-space Sampling}
\label{sec:problem_formulation}
Let $\mathbf{y} \in \mathbb{C}^{M \times N}$ be a complex matrix representing a fully sampled $k$-space, and $\mathbf{x}$ be a reconstruction of $\mathbf{y}$ obtained via Inverse Fourier Transform, $\mathcal{F}^{-1}$; i.e., ${\mathbf x} = \mathcal{F}^{-1}(\mathbf y)$. We simulate partially observed $k$-space by masking $\mathbf{y}$ with a Cartesian binary mask $\mathbf{M}$, resulting in $\mathbf{\tilde{y}}=\mathbf{M}\odot\mathbf{y}$. We denote the corresponding \emph{zero-filled} reconstruction as $\mathbf{\tilde{x}} = \mathcal{F}^{-1}(\mathbf{\tilde{y}})$. In this paper, we use a deep-learning-based reconstruction model $\mathbf{r}(\mathbf{\tilde{x}}; \phi) \rightarrow \mathbf{\hat{x}}$ that takes $\mathbf{\tilde{x}}$ as input, and outputs a de-aliased image reconstruction $\mathbf{\hat{x}}$. The reconstruction model is a convolutional neural network (CNN) parametrized by $\phi$; in particular, we use the network proposed in~\cite{Zhang_2019_CVPR}. Finally, we use subindices to denote time, e.g., $\mathbf{\hat{x}}_{t} = \mathbf{r}({\mathcal{F}^{-1}(\mathbf{M}_t}\odot\mathbf{y}))$ represents the reconstruction obtained at time step $t$ of the acquisition process.
\subsection{Active MRI Acquisition as POMDP}
\label{sec:problem_form}
In our active acquisition formulation, the goal is to learn a \emph{policy} function $\pi(\mathbf{\hat{x}}_t, \mathbf{M}_t; \theta) \rightarrow a_t$ that, at time $t$, maps the tuple of de-aliased image reconstruction~$\mathbf{\hat{x}}_t$, and mask representing observed frequencies~$\mathbf{M}_t$, to the \emph{action}~$a_t$. In Cartesian active MRI acquisition, actions are represented by the unobserved columns of the fully sampled $k$-space, $\mathbf{y}$ (the ones for which the mask $\mathbf{M}_t$ has value of $0$). Once an action is observed, both the mask and the de-aliased image reconstruction are updated to $\mathbf{M}_{t+1}=\mathbf{M}_t+\mathbf{M}^{a_t}$ and $\mathbf{\hat{x}}_{t+1} = \mathbf{r}(\mathcal{F}^{-1}(\mathbf{M}_{t+1} \odot \mathbf{y}))$, where $\mathbf{M}^{a_t}$ is a binary matrix with all zeros except of the column indicated by the action $a_t$. The parameters $\theta$ of the policy are optimized to select a sequence of actions ($k$-space measurements) $[a_t, a_{t+1}, ..., a_T]$ that minimize the acquisition cost function (i.e., maximize the reward function) over all future reconstructions up to time step $T$. In this work, we consider acquisition costs of the form $f(\mathcal{C}(\mathbf{\hat{x}}_1, \mathbf{x}), \mathcal{C}(\mathbf{\hat{x}}_2, \mathbf{x}), ..., \mathcal{C}(\mathbf{\hat{x}}_T, \mathbf{x}))$, where $\mathcal{C}$ represents a pre-defined cost of interest (e.g., Mean Squared Error or negative SSIM), and $f$ is a function that aggregates the costs observed throughout an acquisition trajectory (e.g., sum, area under the metric curve, final metric value at time $T$). Thus, the objective is to minimize the aggregated cost over the \emph{whole range} of MRI acquisition speed-ups. Further, we assume that $f$ can be expressed as a discounted sum of $T$ rewards, one per time step. Under this assumption, we formally introduce the active acquisition problem as a POMDP defined by: 

\begin{compactitem}
    \item \textbf{State set}: The set 
    of all possible tuples $ \mathbf{s}_t \triangleq \langle \mathbf{x}, \mathbf{M}_t \rangle $. Note that the ground truth image, $\bf x$, is hidden and the current mask, $\mathbf{M}_t$, is fully visible.
    \item \textbf{Observation set}: The set %$\mathcal{O}$ 
    of all possible tuples $ \mathbf{o}_t \triangleq \langle \hat{\mathbf{x}}, \mathbf{M}_t \rangle$.
    \item \textbf{Action set}: The set of all possible $k$-space column indices that can be acquired, i.e., $\mathcal{A} \triangleq \{1, 2, ..., W\}$, where $W$ is the image width. Since sampling an already observed $k$-space column does not improve reconstruction, we specify that previously observed columns are invalid actions.
    \item \textbf{Transition function}: Given current mask $\mathbf{M}_t$ and a valid action $a_t \in \mathcal{A}$, the mask component of the state transitions deterministically to $\mathbf{M}_{t+1}=\mathbf{M}_t+\mathbf{M}^{a_t}$, and $\bf x$ remains unchanged. After $T$ steps the system reaches the final time step.
    \item \textbf{Emission function}: In our formulation, we assume that the reconstruction model returns a deterministic reconstruction  $\mathbf{\hat{x}}_t$ at each time step $t$; thus, the observation  after taking action $a_t$ at state $\mathbf{s}_t$ is defined as $\mathbf{o}_t \triangleq \hat{\mathbf{x}}_{t+1}$ (i.e., the reconstruction after adding the new observed column to the mask). %However, in general, the reconstructor could return a probability distribution over reconstructions. 
    \item \textbf{Reward function}: We define the reward as the decrease in reconstruction metric with respect to the previous reconstruction: $R(\mathbf{s}_t, a_t) = \mathcal{C}(\mathbf{\hat{x}}_{t+1}, \mathbf{x}) - \mathcal{C}(\mathbf{\hat{x}}_{t}, \mathbf{x})$. This assumes that $f(\mathcal{C}(\mathbf{\hat{x}}_1, \mathbf{x}), \mathcal{C}(\mathbf{\hat{x}}_2, \mathbf{x}), ..., \mathcal{C}(\mathbf{\hat{x}}_T, \mathbf{x})) \triangleq \mathcal{C}(\mathbf{\hat{x}}_{T}, \mathbf{x})$. We found this reward to be easier to optimize than rewards corresponding to more complex aggregations.
    \item \textbf{Discount factor}: We treat the discount, $\gamma \in [0,1]$, as a hyperparameter.
\end{compactitem}
Note that the above-mentioned POMDP is \emph{episodic}, since the acquisition process has a finite number of steps $T$. At the beginning of each episode, the acquisition system is presented with an initial reconstruction, $\hat{\textbf{x}}_{0}$, of an unobserved ground truth image $\bf x$, as well with an initial subsampling mask $\mathbf{M}_0$. The system then proceeds to iteratively suggest $k$-space columns to sample, and receives updated reconstructions from $\mathbf{r}$. % The process continues for $T$ steps, at which point the episode ends and a new episode starts. 
\subsection{Solving the Active MRI Aquisition POMDP with DDQN}
We start by defining a subject-specific \textsc{ddqn} value network. As mentioned in Section~\ref{sec:background}, POMDP policies are functions of observation histories. However, in active Cartesian MRI acquisition, the whole history of observations is captured by the current observation $\mathbf{o}_t$. Thus, we use $\mathbf{o}_t$ as single input to our value network. We design the value network architecture following~\cite{Zhang_2019_CVPR}'s evaluator network, which receives as input a reconstructed image, $\hat{\mathbf{x}}_t$, and a mask $\mathbf{M}_t$. 
To obtain the reconstructed image $\mathbf{x}_t$ at each time step, we use a pre-trained reconstruction network. % based on~\cite{Zhang_2019_CVPR}'s reconstructor. 
Additionally, we also consider a dataset-specific \textsc{ddqn} variant, which only takes time step information as input (which is equivalent to considering the number of non-zero elements in the mask $\mathbf{M}_t$), and thus, uses the same acquisition trajectory for all subjects in the dataset. 
 
In both cases, we restrict the value network to select among valid actions by setting the value of all previously observed $k$-space columns to $-\infty$. Additionally, we use a modified $\epsilon$-greedy policy~\cite{sutton2018reinforcement} as exploration policy to fill the replay memory buffer. This policy chooses the best action with probability $1-\epsilon$, and chooses an action from the set of valid actions with probability $\epsilon$.

\section{Experimental Results}
\textbf{Datasets and baselines.}
To train and evaluate our models, we use the single-coil knee RAW acquisitions from the fastMRI dataset~\cite{zbontar2018fastMRI}, composed of 536 public training set volumes and 97 public validation set volumes. We create a held-out test set by randomly splitting the public validation set into a new validation set with 48 volumes, and a test set with 49. This results in \numc{19878} 2D images for training, \numc{1785} images for validation, and \numc{1851} for test. All data points are composed of a $640 \times 368$ complex valued $k$-space matrix with the $36$ highest frequencies zero-padded. In all experiments, we use vertical Cartesian masks, such that one action represent acquisition of one column in the $k$-space matrix. Hence, the total number of possible actions for this dataset is $332$. 

We compare our approach to the following heuristics and baselines: (1) \underline{Random policy} (\textsc{random}): Randomly select an unobserved $k$-space column; (2) \underline{Random with low frequency bias} (\textsc{random-lb}): Randomly select an unobserved $k$-space column, favoring low frequency columns; (3) \underline{Low-to-high policy} (\textsc{lowToHigh}): Select an unobserved $k$-space column following a low to high frequency order; and (4) \underline{Evaluator policy} (\textsc{evaluator}): Select an unobserved $k$-space column following the observation scoring function introduced by~\cite{Zhang_2019_CVPR}. To ensure fair comparisons among all methods, we fix same number of low-frequency observations (for details see Section~\ref{sec:details}). Moreover, for reference, we also include results for a one-step oracle policy that, having access to ground truth at test time, chooses the frequency that will reduce $\mathcal{C}$ the most (denoted as \textsc{oracle}).
\iffalse
\begin{compactitem}
    \item \underline{Random policy} (\textsc{random}): Randomly select an unobserved $k$-space column.
    \item \underline{Random with low frequency bias} (\textsc{random-lb}): Randomly select an unobserved $k$-space column, favoring low frequency columns.
    \item \underline{Low-to-high policy} (\textsc{lowToHigh}): Select an unobserved $k$-space column following a low to high frequency order.
    \item \underline{Evaluator policy} (\textsc{evaluator}): Select an unobserved $k$-space column following the observation scoring function introduced by~\cite{Zhang_2019_CVPR}.
\end{compactitem}
\fi
\begin{figure}[t]
    \centering
    \subfloat[Scenario-2L ]{\includegraphics[width=0.50\columnwidth]{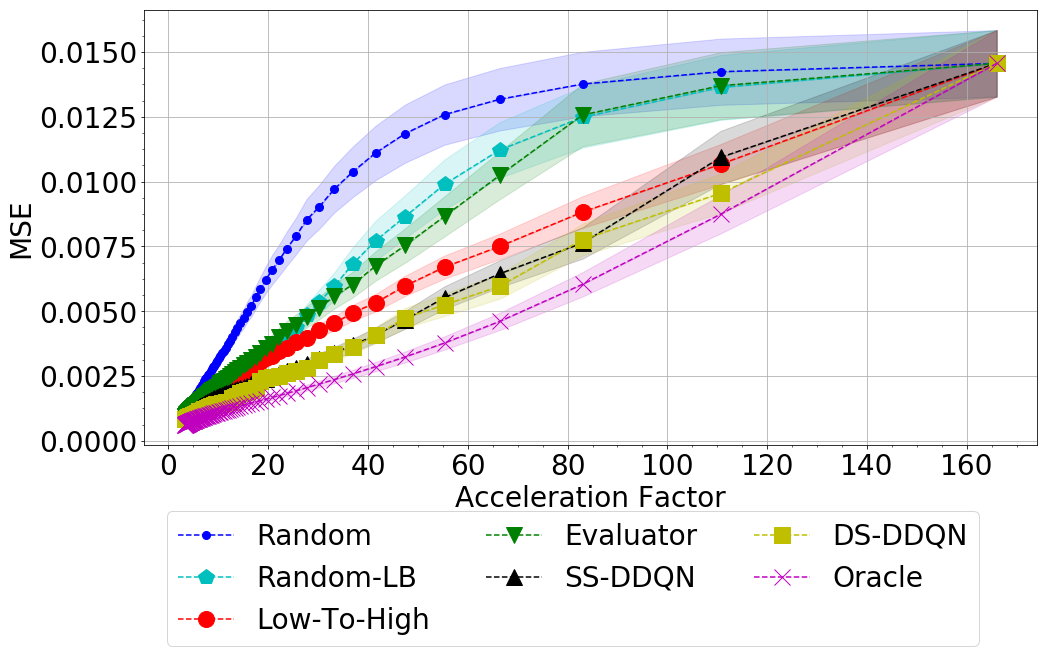}}
    \hfill
    \subfloat[Scenario-30L]{\includegraphics[width=0.50\columnwidth]{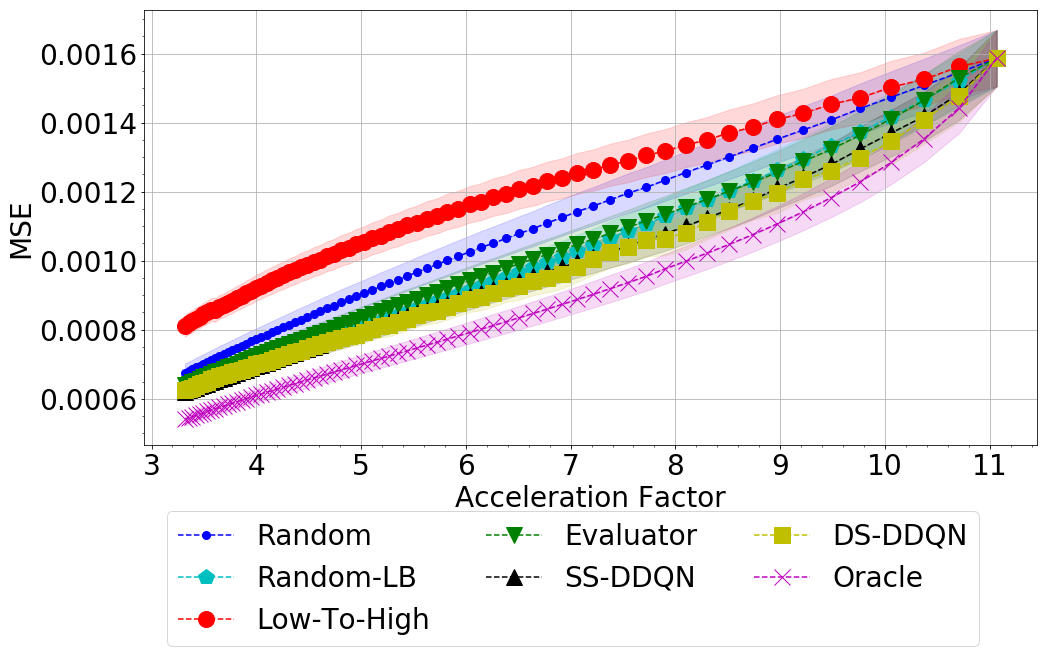}}

    \caption{Reconstruction MSE vs. acceleration factor for different strategies. Both \textsc{ddqn} models outperform all baselines and heuristics for the vast majority of acceleration factors considered.}
    \label{fig:mse_vs_accel}
\end{figure}

\textbf{Training details.}
\label{sec:details}
In our experiments, we use the reconstruction architecture from \cite{Zhang_2019_CVPR} and train it using negative log-likelihood on the training set. Following \cite{Zhang_2019_CVPR}, the reconstructor is trained on the whole range of acceleration factors, by randomly sampling masks with different patterns and acceleration factors each time an image is loaded. Similarly to \cite{Zhang_2019_CVPR}, we also force a fixed number of low frequencies to be always observed, and train two versions of the reconstruction model: one that always observes 30 low frequencies---referred to as Scenario-30L---, corresponding to $\sim3\!\!\times-11\times$ accelerations, and another that always observes 2 low frequencies---referred to as Scenario-2L---, corresponding to extreme acceleration factors of up to $166\times$ acceleration. In contrast to \cite{Zhang_2019_CVPR}, we train the reconstructor and the policy networks in stages, as we noticed that it leads to better results. Note that none of the architectures used in the experiments uses complex convolutions.
% in our experiments we refer to these two cases as Scenario-30L and Scenario-2L, respectively. Additionally, in contrast to \cite{Zhang_2019_CVPR}, we train the reconstructor and the acquisition functions in stages 
% (e.g., first we train reconstructor to convergence and then we train the acquisition function) 
%
For both \textsc{ddqn} approaches, we experimented with the reward defined in Section~\ref{sec:problem_form} and four choices of $\mathcal{C}$: Mean Squared Error (MSE), Normalized MSE, negative Peak Signal to Noise Ratio (PSNR) and negative Structural Similarity (SSIM). For each metric, we trained a value network using a budget of $T=100 - L$ actions, where $L=2$ or $L=30$ is the number of initial fixed low frequencies, a discount factor $\gamma=0.5$, and a replay buffer of \numc{20000} tuples. Each training episode starts with a mask with a fixed number of low frequencies $L$. %; we consider $L=2$ and $L=30$, each trained with its corresponding reconstructor.
The \textsc{ddqn} models are trained for \numc{5000000} state transition steps, roughly equivalent to 3.6 iterations over the full training set for $L=30$, or 2.6 for $L=2$. We periodically evaluate the current greedy policy on the validation set, 
%on a separate thread running continuously, 
and use the best validation result to select the final model for testing. To reduce computation, we used a random subset of \numc{200} validation images when training \textsc{ddqn}. 
% Finally, we consider a \textsc{ddqn} model that only takes time step information as input (which is equivalent to only the number of non-zero elements in the mask $\mathbf{M}_t$).
% We denote by \textsc{ddqn-ps} the version that uses the the full observation, and by % This model produces a deterministic policy of time that results in the same acquisition order for all images in the dataset. 
\begingroup
\setlength{\tabcolsep}{3pt}
\begin{table}[t]
\resizebox{\linewidth}{!}{% 
    \centering
    % \begin{tabular}{|c|c|c|c|c|c|c|}
    \begin{tabular}{lccccccc}
    \hline
        Metric & \textsc{random} & \textsc{random-lb} & \textsc{lowToHigh} & \textsc{evaluator} & \textsc{ss-ddqn} & \textsc{ds-ddqn} \\ \hline
         MSE ($\times 10^{-3}$) $\downarrow$ & 8.90 (0.41) & 8.24 (0.37) & 9.64 (0.45) & 8.33 (0.38) & 8.00 (0.35) & \textbf{7.94 (0.35)} \\ %\hline
         NMSE ($\times 10^{-1}$) $\downarrow$ & 3.02 (0.16) & 2.93 (0.16) & 3.13 (0.17) & 3.06 (0.17) & 2.88 (0.17) & \textbf{2.87 (0.16)} \\ %\hline
         PSNR ($\times 10^2$) $\uparrow$ & 2.23 (1.28) & 2.25 (1.75) & 2.21 (1.23) & 2.24 (1.33) & \textbf{2.27 (1.34)} & 2.26 (1.35) \\ %\hline
         SSIM  $\uparrow$ & 4.77 (0.06) & 4.82 (0.07) & 4.71 (0.06) & 4.78 (0.07) & \textbf{4.86 (0.07)} & \textbf{4.86 (0.07)} \\ \hline
    \end{tabular}
    }
    \caption{Average test set AUC with 95\% confidence intervals (one AUC value/image) for 6 different active acquisition policies (Scenario-30L).}
    % \vspace{-18pt}
    \label{tab:auc_15_lines}
\end{table}
\endgroup
\begingroup
\setlength{\tabcolsep}{3pt}
\begin{table}[t]
\resizebox{\linewidth}{!}{% 
    \centering
    \begin{tabular}{lccccccc}
    \hline
        Metric & \textsc{random} & \textsc{random-lb} & \textsc{lowToHigh} & \textsc{evaluator} & \textsc{ss-ddqn} & \textsc{ds-ddqn} \\ \hline
         MSE $\downarrow$ & 1.97 (0.17) & 1.73 (0.15) & 1.39 (0.10) & 1.70 (0.15) & 1.31 (0.11) & \textbf{1.24 (0.10)}  \\ %\hline
         NMSE $\downarrow$ & 20.6 (0.43) & 18.7 (0.41) & 17.5 (0.37) & 17.7 (0.42) & 15.5 (0.35) & \textbf{15.1 (0.34)} \\ %\hline
         PSNR ($\times 10^3$) $\uparrow$ & 3.63 (0.02) & 3.73 (0.02) & 3.76 (0.02) & 3.79 (0.02) & 3.89 (0.02) & \textbf{3.90 (0.02)}  \\ %\hline
         SSIM $\uparrow$ & 73.2 (1.19) & 75.3 (1.22) & 73.3 (1.19) & 76.0 (1.23) & \textbf{78.0 (1.25)} & \textbf{78.0 (1.25)} \\ \hline
    \end{tabular}
    }
    \caption{Average test set AUC with 95\% confidence intervals (one AUC value/image) for 6 different active acquisition policies (Scenario-2L).}
    \label{tab:auc_1_line}
\end{table}
\endgroup

\subsection{Results}
Fig.~\ref{fig:mse_vs_accel} (a-b) depicts the average test set MSE as a function of acceleration factor, with Scenario-2L on Fig.~\ref{fig:mse_vs_accel} (a) and Scenario-30L on Fig.~\ref{fig:mse_vs_accel} (b).\footnote{Plots for NMSE, SSIM and PSNR are available in the supplementary material.} Results show that both \textsc{ddqn} models outperform all considered heuristics and baselines for the vast majority of acceleration factors. In Scenario-30L, the mean MSE obtained with our models is between 3-7\% lower than the best baseline (\textsc{evaluator}), for all accelerations between $4\times$ and $10\times$. For the case of extreme accelerations (Scenario-2L), our best model (dataset-specific \textsc{ddqn}) outperforms the best heuristic (\textsc{lowToHigh}) by at least 10\% (and up to 35\%) on all accelerations below $100\times$. Note that for this scenario and the MSE metric, the performance of \textsc{random} and \textsc{evaluator} deteriorated significantly compared to Scenario-30L.
To further facilitate comparison among all methods, we summarize the overall performance of an acquisition policy into a single number, by estimating, for each image, the area under curve (AUC), and averaging the resulting values over the test set. Results are summarized in Tab.~\ref{tab:auc_15_lines} for Scenario-30L and Tab.~\ref{tab:auc_1_line} for Scenario-2L. In all cases, the \textsc{ddqn} policies outperform all other models. In Scenario-30L, the improvements of our models over the best baselines range from 0.55\% to 2.9\%, depending on the metric. In Scenario-2L, the improvements range from 2.68\% to 11.6\%. Further, paired t-tests (pairing over images) between the AUCs obtained with our models and those of the best baseline, for each metric, indicate highly significant differences, with $p$-values generally lower than $10^{-8}$. 
Interestingly, we found that the data-specific \textsc{ddqn} slightly outperforms the subject-specific one for most metrics. While this seems to suggest that subject-specific trajectories are not necessary, we point out that the gap between the performance of \textsc{oracle} and the models considered indicates the opposite. In particular, policy visualizations for \textsc{oracle} (see Section \ref{sec:policies}) show wide subject-specific variety. Therefore, we hypothesize that the better performance of \textsc{ds-ddqn} is due to optimization and learning stability issues.  

\begin{figure}[t]
    \centering
    
    \subfloat[Scenario-2L]{\includegraphics[width=0.5\columnwidth]{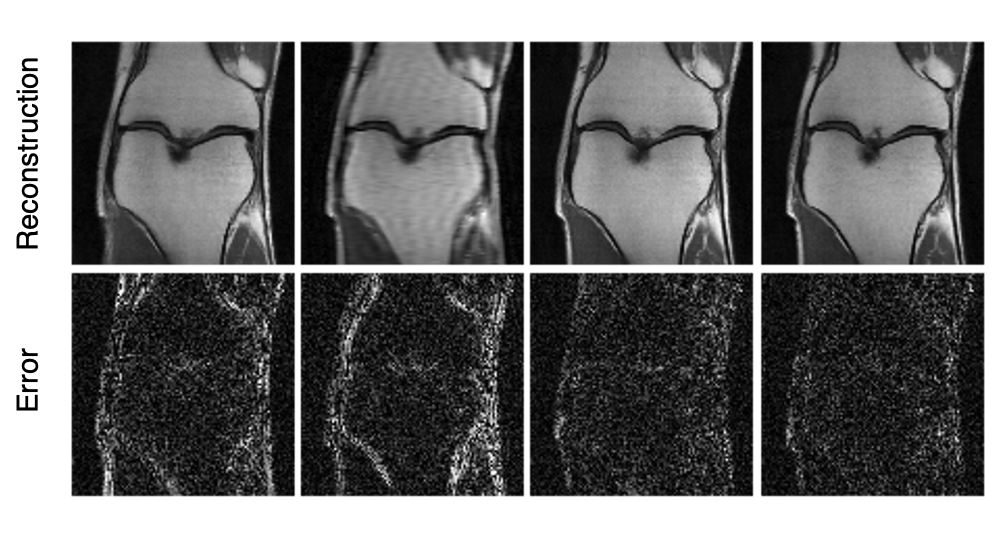}}
    \subfloat[Scenario-30L]{\includegraphics[width=0.5\columnwidth]{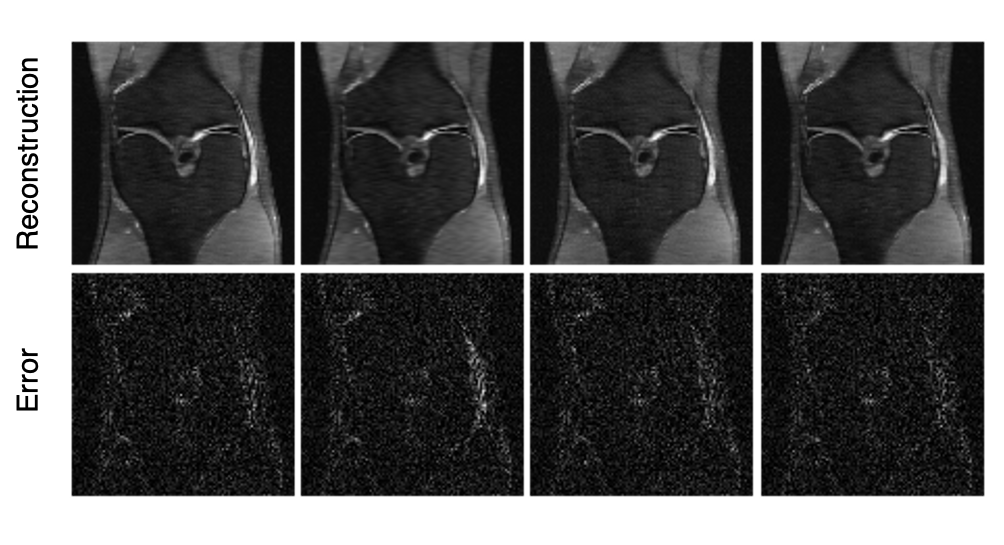}}
    \caption{Reconstructions and error maps under 4 acquisition policies (from left:\textsc{random}, \textsc{lowToHigh}, \textsc{evaluator}, \textsc{ss-ddqn}) at $10\times$ acceleration for Scenario-2L and $8\times$ for Scenario-30L. The images depict magnitude information. Note that the subfigures (a) and (b) depict different knee images. Additional images are shown in the supplementary material.}
    \label{fig:example1}
\end{figure}

\textbf{Qualitative results.} Fig.~\ref{fig:example1} shows examples of reconstructions together with error maps for four different acquisition policies (\textsc{random}, \textsc{lowToHigh}, \textsc{evaluator}, subject-specific \textsc{ddqn}). We display the $10\times$ and $8\times$ acceleration for Scenario-2L and Scenario-30L, respectively. Looking at the error maps (bottom row), the differences in reconstruction quality between the subject-specific \textsc{ddqn} and the rest of the policies are substantial, and the reconstruction is visibly sharper and more detailed than the ones obtained with the baselines.

\subsection{Policy Visualization}
\label{sec:policies}
\begin{figure}[t]
    \centering
    \subfloat[\textsc{ss-ddqn}$_{mse}$]{\includegraphics[width=0.2\columnwidth]{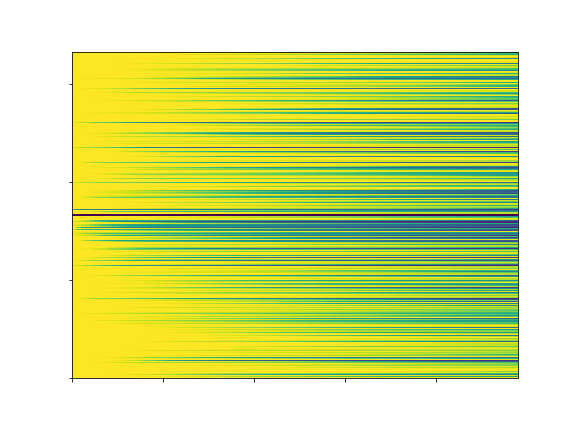}}
    \subfloat[\textsc{ds-ddqn}$_{mse}$]{\includegraphics[width=0.2\columnwidth]{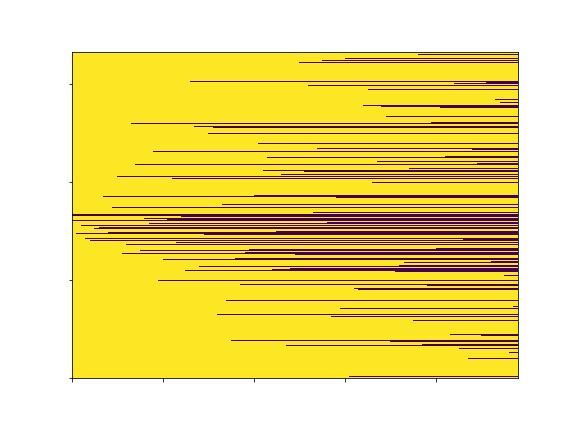}}
    \subfloat[\textsc{ss-ddqn}$_{ssim}$]{\includegraphics[width=0.2\columnwidth]{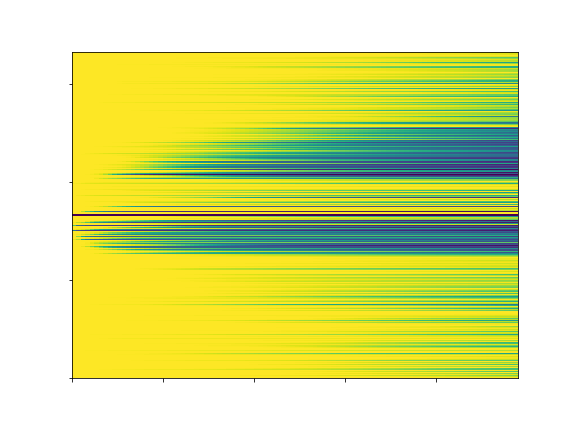}}
    \subfloat[\textsc{ds-ddqn}$_{ssim}$]{\includegraphics[width=0.2\columnwidth]{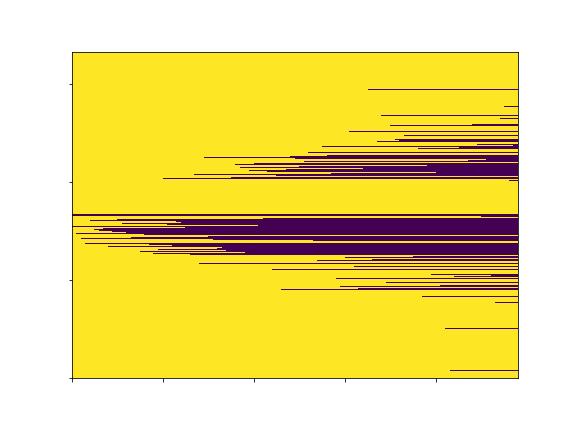}}
    \subfloat[\textsc{oracle}]{\includegraphics[width=0.2\columnwidth]{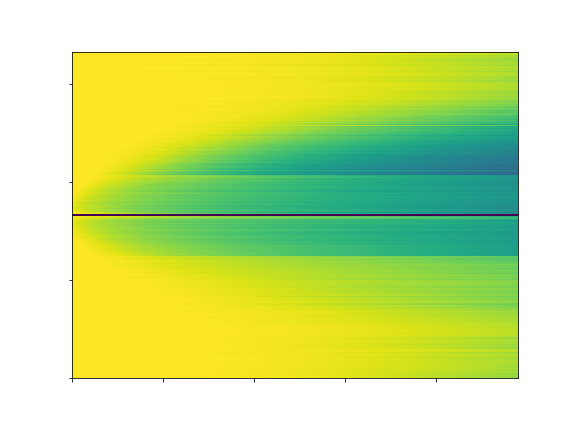}}
    \newline
    \subfloat[\textsc{ss-ddqn}$_{mse}$]{\includegraphics[width=0.2\columnwidth]{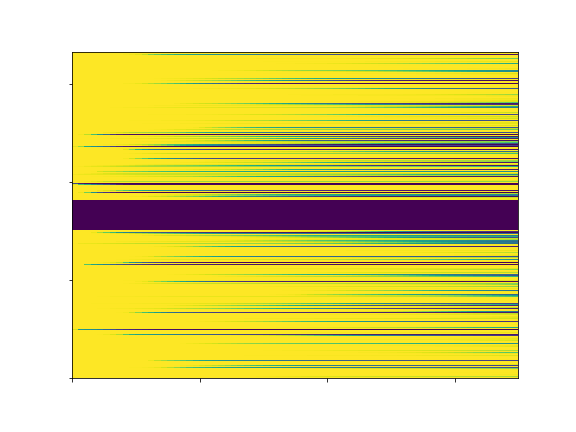}}
    \subfloat[\textsc{ds-ddqn}$_{mse}$]{\includegraphics[width=0.2\columnwidth]{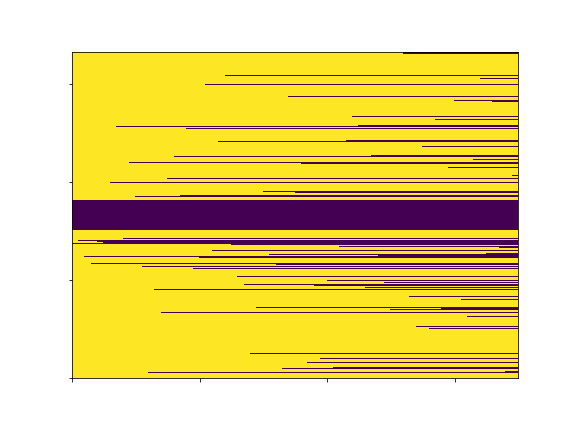}}
    \subfloat[\textsc{ss-ddqn}$_{ssim}$]{\includegraphics[width=0.2\columnwidth]{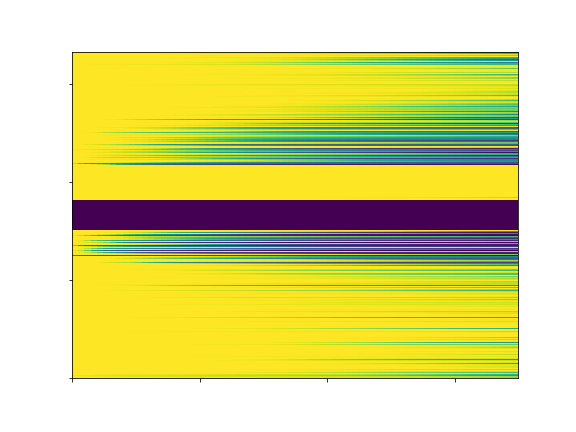}}
    \subfloat[\textsc{ds-ddqn}$_{ssim}$]{\includegraphics[width=0.2\columnwidth]{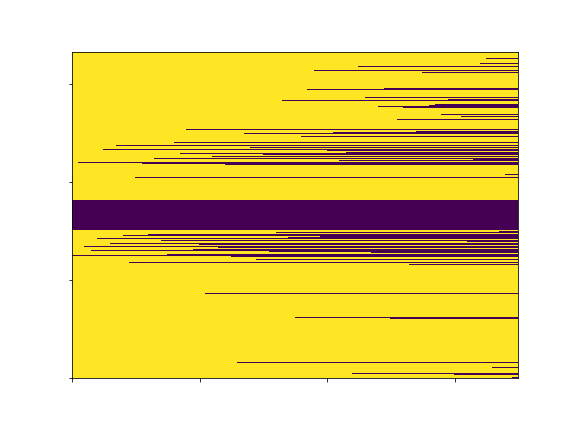}}
    \subfloat[\textsc{oracle}]{\includegraphics[width=0.2\columnwidth]{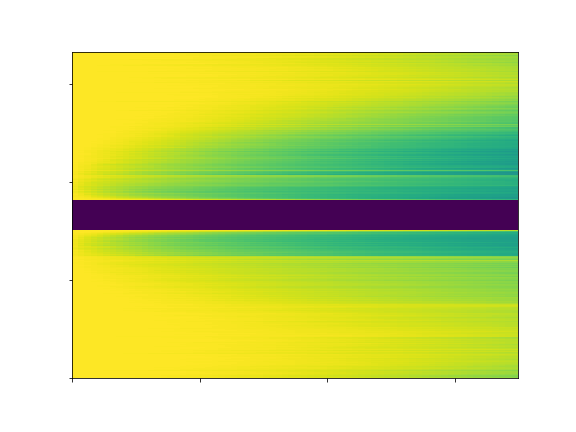}}
     \newline
    \caption{Policy visualizations for \textsc{ddqn} models: Scenario-2L (a-e) and Scenario-30L (e-j). Subfigures (e) and (j) shows oracle policy obtained with SSIM criteria. See main text for details. See supplementary for additional results.}
    \label{fig:policies-2l}
\end{figure}

Fig.~\ref{fig:policies-2l} illustrates \textsc{ddqn} policies on the test set for \textsc{ddqn} trained with MSE and negative SSIM acquisition costs. Each row in the heat maps represents a cumulative distribution function of the time ($x$-axis) at which the corresponding $k$-space frequency is chosen ($y$-axis); the darker color the higher the values. Note that low frequencies are closer to the center of the heat map, while high frequencies are closer to the edges. In the dataset-specific \textsc{ddqn} heat maps, each row instantly transitions from light to dark intensities at the time where the frequency is chosen by the policy (recall that this policy is a deterministic function of time). In the subject-specific \textsc{ddqn}, smoother transitions indicate that frequencies are not always chosen at the same time step. Furthermore, one can notice that some frequencies are more likely to be chosen earlier, indicated by a dark intensity appearing closer to the left side of the plot. Overall, for both models and costs, we observe a tendency to start the acquisition process by choosing low and middle frequencies, while incorporating high frequencies relatively early in the process. However, when comparing MSE-based to SSIM-based policies, we observe that the SSIM policy is more biased towards low frequencies and it seems to take advantage of $k$-space Hermitian symmetry -- only few actions selected in the upper-center part of the SSIM heat maps.

\section{Conclusion}
In this paper, we formulated the active MRI acquisition problem as a Partially Observable Markov Decision Process and solved it using the Double Deep Q-Network algorithm. On a large scale single-coil knee dataset, we learned policies that outperform, in terms of four metrics (MSE, NMSE, SSIM and PSNR), simple acquisition heuristics and the scoring function introduced in~\cite{Zhang_2019_CVPR}. We also observed that the dataset-specific \textsc{ddqn} slightly outperforms the subject-specific \textsc{ddqn}, and that a gap still exists between our models and the best possible performance (illustrated by an oracle policy). This performance gap encourages further research to improve models and algorithms to address the active MRI acquisition problem. Finally, it is important to note that our experimental setup is simplified for the purpose of model exploration (e.g. we do not consider all the practical MRI phase-encoding sampling issues).

\textbf{Acknowledgements.} The authors would like to thank the fastMRI team at FAIR and at NYU for engaging discussions. We would like to express our gratitude to Amy Zhang and Joelle Pineau for helpful pointers and to Matthew Muckley for providing feedback on an early draft of this work.

We also acknowledge the Python community for developing the core set of tools that enabled this work, including PyTorch~\cite{paszke2017automatic}, Numpy~\cite{oliphant2006guide}, Jupyter~\cite{kluyver2016jupyter}, scikit-learn~\cite{pedregosa2011scikit}, scikit-image~\cite{van2014scikit}, and Matplotlib~\cite{hunter2007matplotlib}.
% \newpage
{\small
\bibliographystyle{ieee_fullname}
\bibliography{pineda-et-al-20}
}
\clearpage
\section*{Supplementary}
% The supplementary material contains additional results and visualization: (1) Fig.~\ref{fig:ssim_vs_accel} contains the plots of NMSE, PSNR and SSIM of the test set as a function of acceleration for both tested scenarios (Scenario-2L and Scenario-30L); (2) Fig.~\ref{fig:policies-suppl} depicts policy visualizations for all heuristics and baselines ans well as \textsc{ddqn} trained with NMSE and PSNR; and (3) Fig.~\ref{fig:example_full} shows additional qualitative results for different accelerations.

\setcounter{page}{1}
    
\begin{figure}[htbp]
    \centering
    \subfloat[SSIM: Scenario-2L ]{\includegraphics[width=0.42\columnwidth]{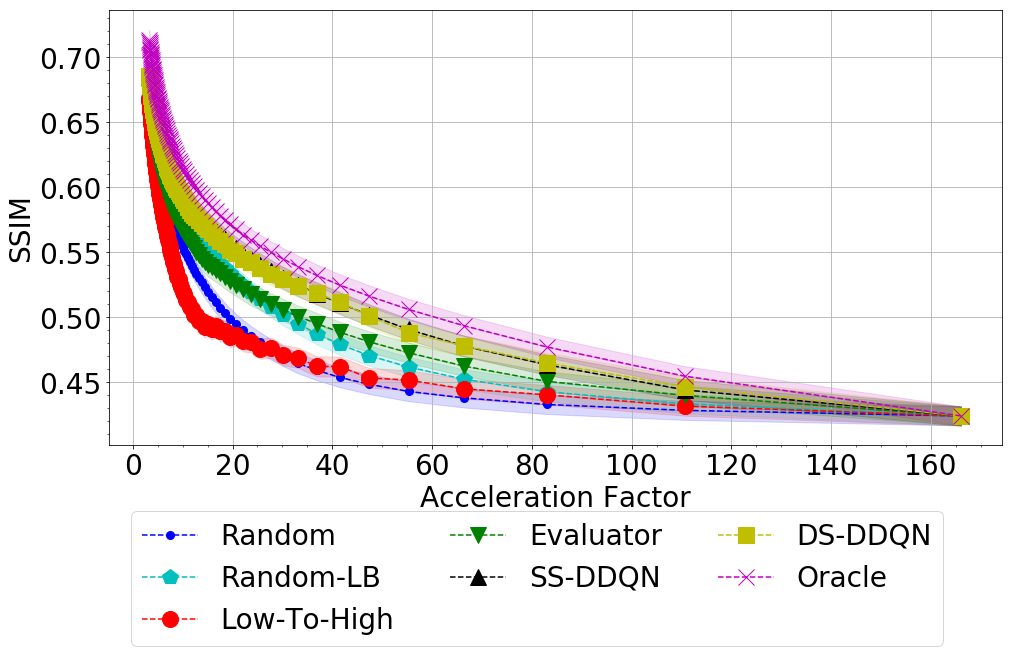}}
    \hfill
    \subfloat[SSIM: Scenario-30L]{\includegraphics[width=0.42\columnwidth]{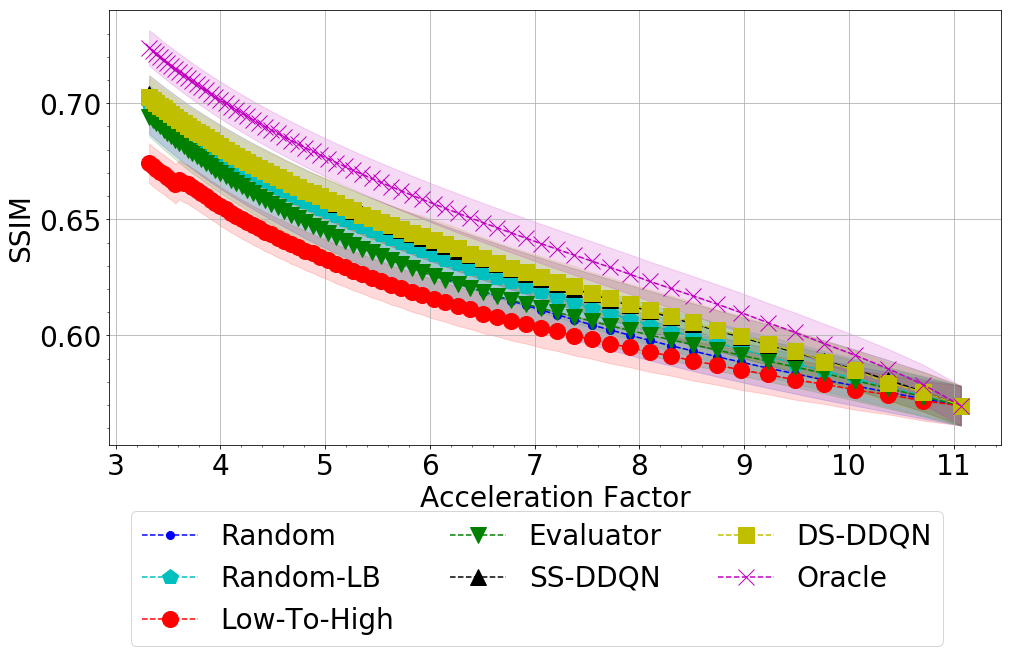}}
    \\
    \subfloat[PSNR: Scenario-2L ]{\includegraphics[width=0.42\columnwidth]{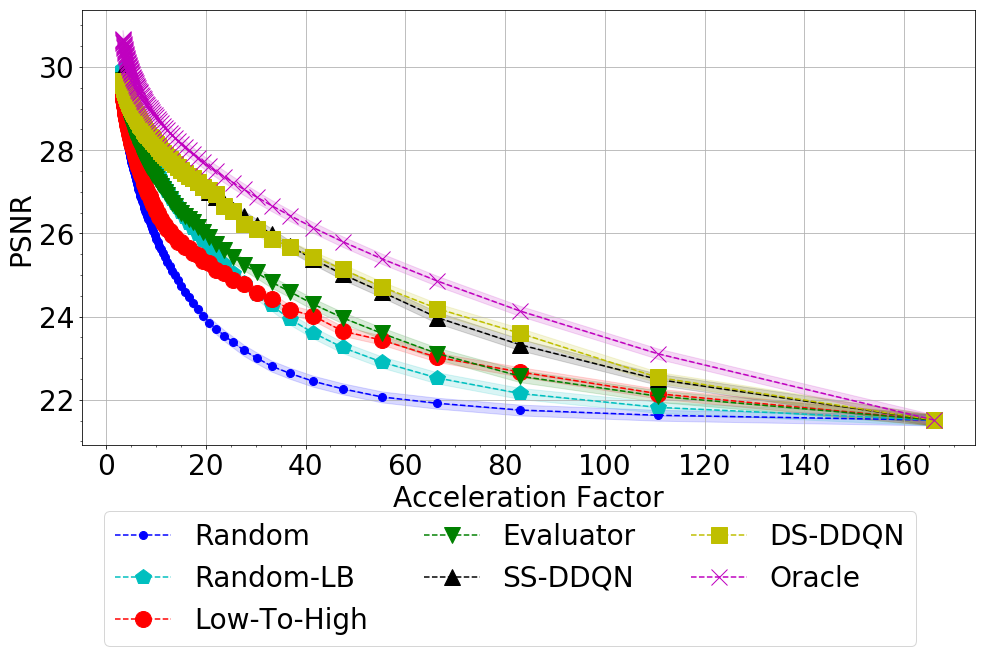}}
    \hfill
    \subfloat[PSNR: Scenario-30L]{\includegraphics[width=0.42\columnwidth]{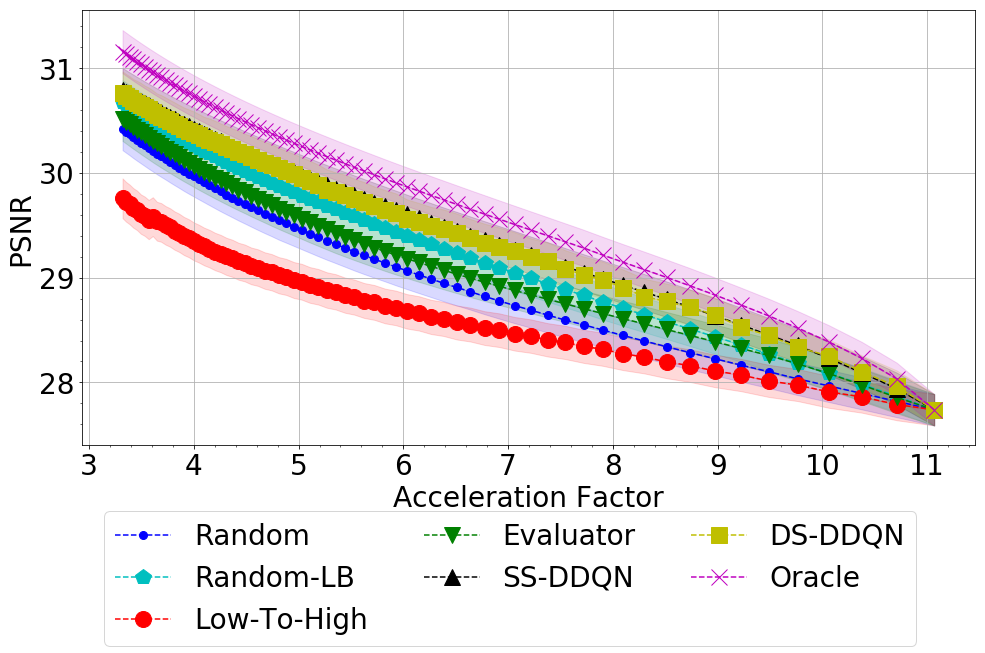}}
    \\
    \subfloat[NMSE: Scenario-2L ]{\includegraphics[width=0.42\columnwidth]{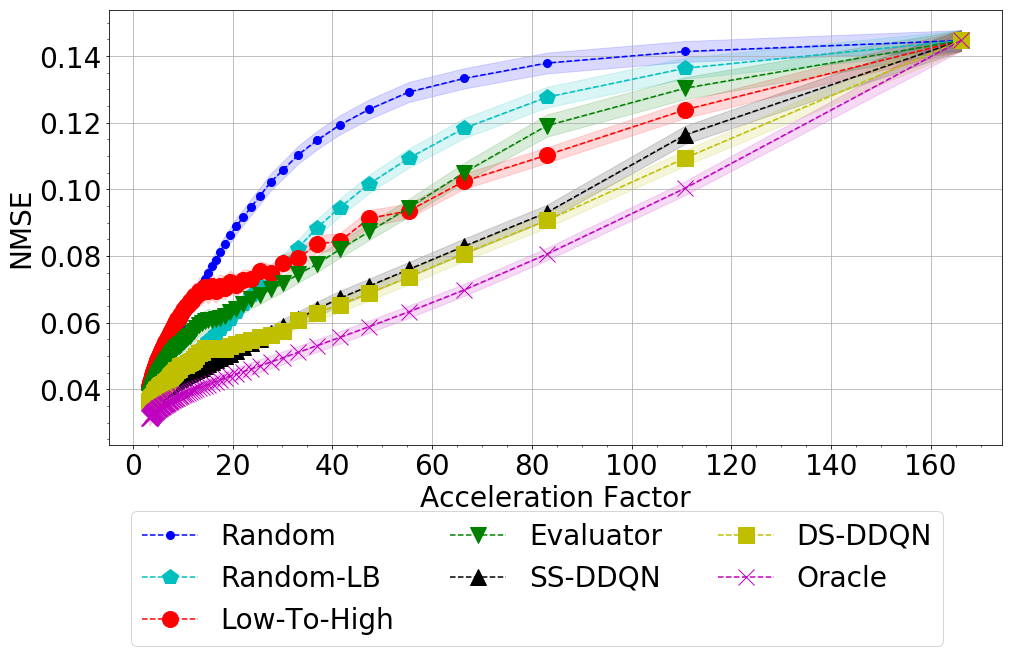}}
    \hfill
    \subfloat[NMSE: Scenario-30L]{\includegraphics[width=0.42\columnwidth]{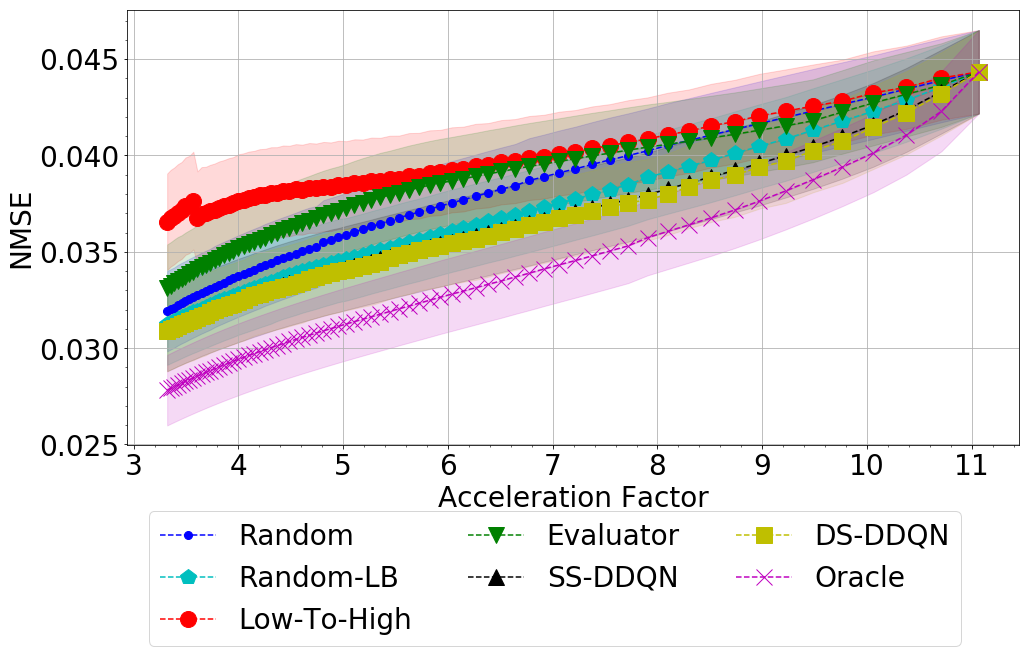}}
    \\
    \caption{Reconstruction quality vs. acceleration factor for different acquisition strategies. Our \textsc{ddqn} models outperform all baselines for most acceleration factors in all metrics. \textsc{ddqn} models are trained with rewards based on the corresponding metric. For MSE metric, see the main body of the paper. }
    \vspace{-12pt}
    \label{fig:ssim_vs_accel}
\end{figure}

\begin{figure}[h]
    \centering
    \subfloat[\textsc{random}]{\includegraphics[width=0.24\columnwidth]{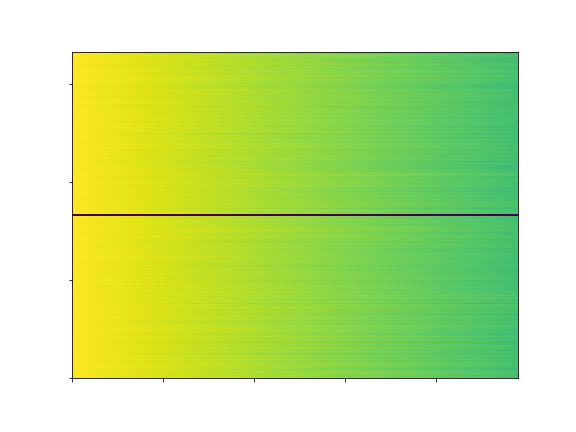}} \hfill
    \subfloat[\textsc{random-lb}]{\includegraphics[width=0.24\columnwidth]{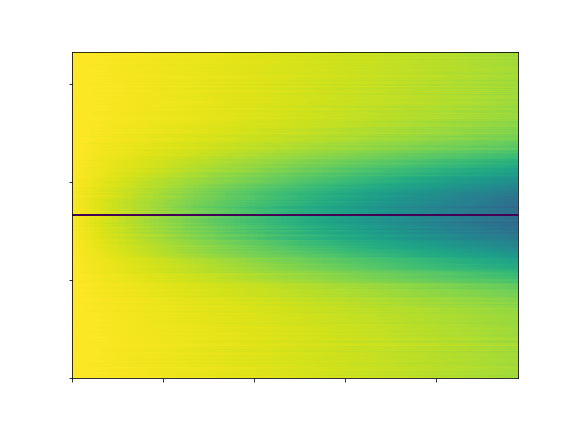}} \hfill
    \subfloat[\textsc{lowToHigh}]{\includegraphics[width=0.24\columnwidth]{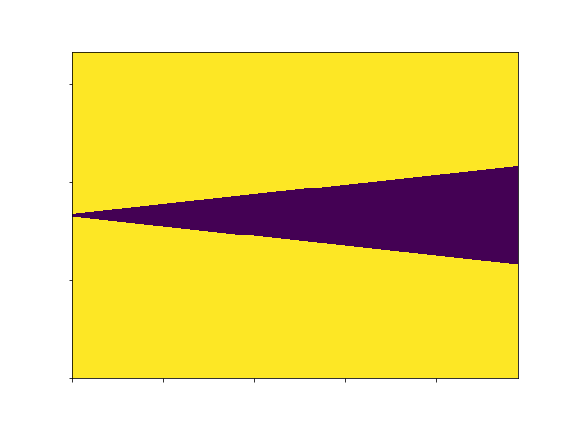}} \hfill
    \subfloat[\textsc{evaluator}]{\includegraphics[width=0.24\columnwidth]{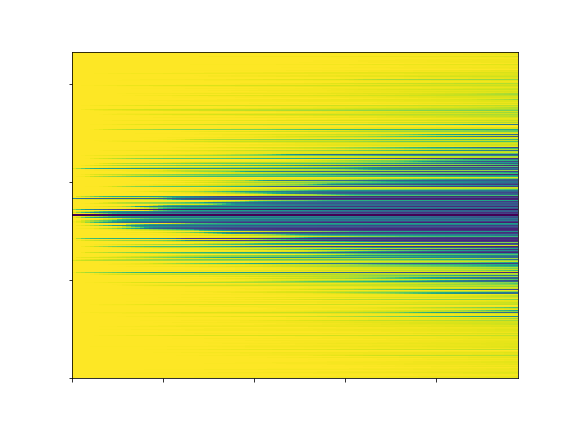}} \hfill
    % \subfloat[\textsc{oracle}]{\includegraphics[width=0.2\columnwidth]{figures/policies_viridis/oracle_1.png}}
    \newline
    \subfloat[\textsc{ss-ddqn}$_{nmse}$]{\includegraphics[width=0.24\columnwidth]{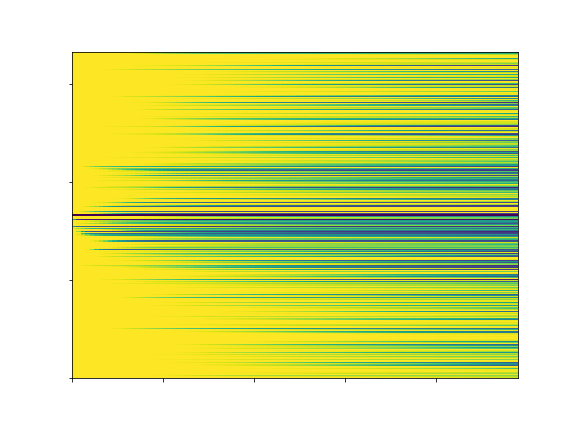}} \hfill
    \subfloat[\textsc{ds-ddqn}$_{nmse}$]{\includegraphics[width=0.24\columnwidth]{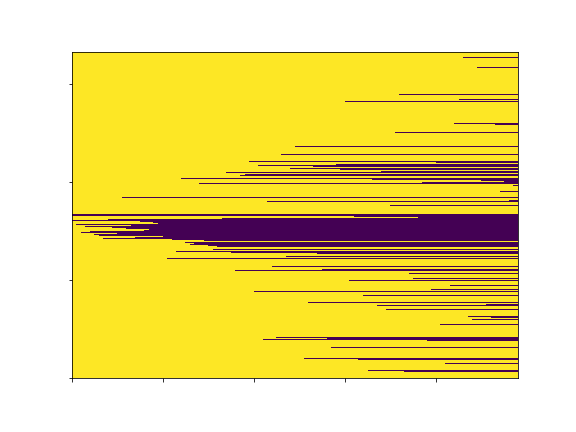}} \hfill
    \subfloat[\textsc{ss-ddqn}$_{psnr}$]{\includegraphics[width=0.24\columnwidth]{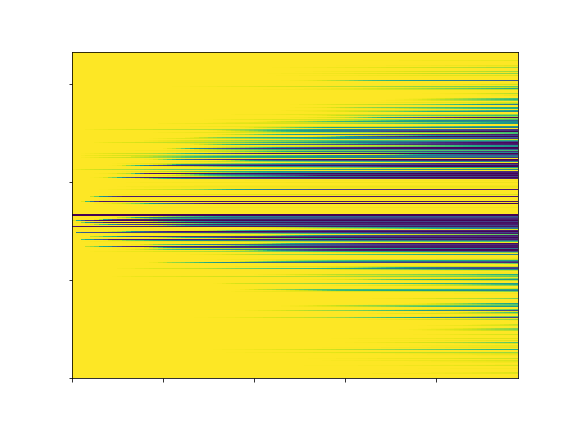}} \hfill
    \subfloat[\textsc{ds-ddqn}$_{psnr}$]{\includegraphics[width=0.24\columnwidth]{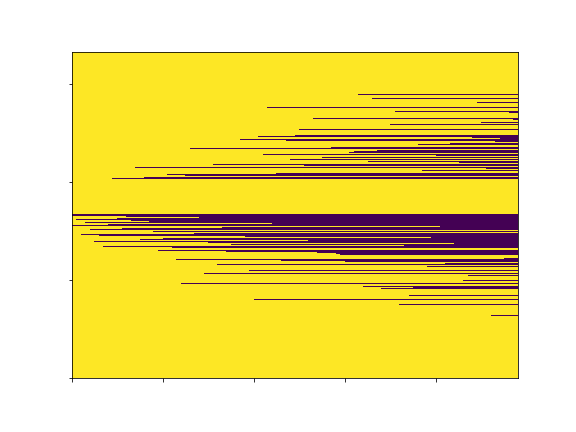}}
    \newline
    % \subfloat[\textsc{random}]{\includegraphics[width=0.2\columnwidth]{figures/policies_viridis/random_15.png}}
    % \subfloat[\textsc{random-lb}]{\includegraphics[width=0.2\columnwidth]{figures/policies_viridis/random-lb_15.png}}
    % \subfloat[\textsc{lowToHigh}]{\includegraphics[width=0.2\columnwidth]{figures/policies_viridis/low-to-high_15.png}}
    % \subfloat[\textsc{evaluator}]{\includegraphics[width=0.2\columnwidth]{figures/policies_viridis/evaluator_15.png}}
    % \subfloat[\textsc{oracle}]{\includegraphics[width=0.2\columnwidth]{figures/policies_viridis/oracle_15.png}}
    
    \caption{Policy visualizations for all heuristics and baselines for  Scenario-2L (a-d) as well as \textsc{ddqn} policies trained with NMSE and PSNR. For \textsc{ddqn} policies trained with MSE and SSIM, see the main body of the paper.}
    \label{fig:policies-suppl}
\end{figure}

\begin{figure}[t]
    \centering
    
    \includegraphics[width=0.98\columnwidth]{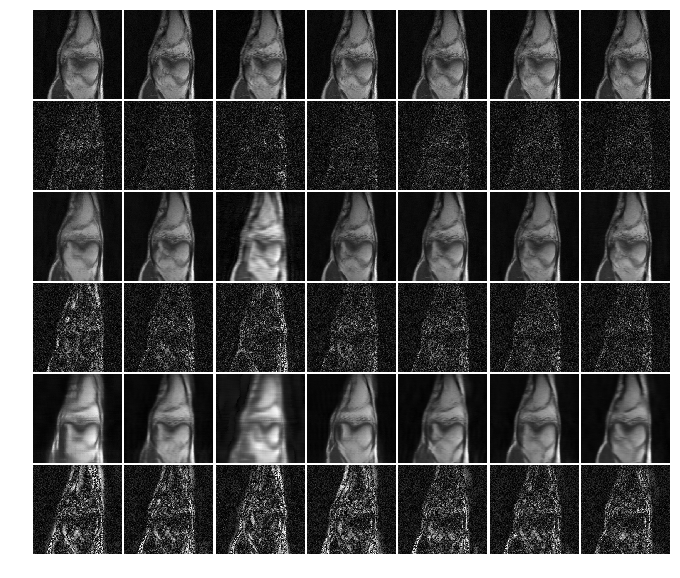}
    
    \caption{Example of image reconstructions and error maps under all different acquisition policies (Scenario-2L). Top to bottom: 4X, 16X, 64X acceleration. Left to right: \textsc{random}, \textsc{random-lb}, \textsc{lowToHigh}, \textsc{evaluator}, \textsc{ss-ddqn}, \textsc{ds-ddqn}, \textsc{oracle}.}
    \label{fig:example_full}
\end{figure}

\end{document}